\documentclass[preprint,prd,noshowpacs,nofootinbib]{revtex4}

\usepackage{color} 
\usepackage{amssymb}
\usepackage{amsmath}
\usepackage{graphicx} 
\usepackage{float}
\usepackage{braket}     
\usepackage{pdfpages}
\usepackage{epstopdf}
\usepackage[utf8]{inputenc}
\usepackage{comment}
\usepackage{xspace}

\DeclareUnicodeCharacter{202F}{\,}

\begin{document}
 
\title{Bubble Universe from Flat Spaces}

\author{Eduardo Guendelman}
\email{guendel@bgu.ac.il} 
\affiliation{Department of Physics, Ben-Gurion University of the Negev, Beer-Sheva, Israel \vspace{0.15in} \\
Frankfurt Institute for Advanced Studies, Giersch Science Center, Campus Riedberg, Frankfurt am Main, Germany \vspace{0.15in} \\
Bahamas Advanced Studies Institute and Conferences, 4A Ocean Heights, Hill View Circle, Stella Maris, Long Island, The Bahamas}

\author{Jacov Portnoy}
\email{jacovportnoy@gmail.com}
\affiliation{Bahamas Advanced Studies Institute and Conferences, 4A Ocean Heights, Hill View Circle, Stella Maris, Long Island, The Bahamas}

\date{\today}

\begin{abstract}
We show by matching two flat spaces one in Minkowski coordinates ( empty space) and the other in Minkowski coordinates after a special conformal transformation (also empty space) through a bubble with positive and constant surface tension, that the motion of the bubble is hyperbolic. If the surface tension is very big the initial size of the bubble is as small as we wish, so that we can indeed obtain an infinite universe out of empty spaces. The induced space in the bubble is de Sitter type.
\end{abstract}

\maketitle

\email{guendel@bgu.ac.il}
\affiliation{Department of Physics, Ben-Gurion University of the Negev, Beer-Sheva, Israel.\\}
\affiliation{Frankfurt Institute for Advanced Studies (FIAS),
Ruth-Moufang-Strasse 1, 60438 Frankfurt am Main, Germany.\\}
\affiliation{Bahamas Advanced Study Institute and Conferences, 
4A Ocean Heights, Hill View Circle, Stella Maris, Long Island, The Bahamas.}

\section{Introduction}
There are many interesting situations where space time is flat everywhere, except for some lines or surfaces, may be the most famous example being  the cosmic string  \cite{Cosmic Strings}, where space is flat everywhere except on a line, the string, where a conical singularity resides. 

Naturally one may ask if instead of a line, we may have a surface matching two flat spaces, expressed nevertheless in different coordinate systems. This seems interesting and fundamental because it may shed light into the vacuum structure of the theory, and finding novel ways to connect different vacuum states. The resulting solutions also hint to new ways to generate universes from nothing when continued to euclidean space.

We will study here the matching of Flat Minkowski space and Flat Minkowski space after a special conformal Transformation and then compare with results obtained from considerations of string matter and of braneworlds in the context of string theories with dynamical tension.
\section{Flat Minkowski space and Flat Minkowski space after a special conformal Transformation}

The flat spacetime in Minkowski coordinates is,

 \begin{equation}\label{Minkowski}
 ds_1^2 = \eta_{\alpha \beta} dx^{\alpha} dx^{\beta}
\end{equation}

where $ \eta_{\alpha \beta}$ is the standard Minkowski metric, with 
$ \eta_{00}= 1$, $ \eta_{0i}= 0 $ and $ \eta_{ij}= - \delta_{ij}$.
This is of course a solution of the vacuum Einstein´s equations.

We now consider the conformally transformed metric

 \begin{equation}\label{Conformally transformed Minkowski}
 ds_2^2 = \Omega(x)^2  \eta_{\alpha \beta} dx^{\alpha} dx^{\beta}
\end{equation}
where the conformal factor coincides with that obtained from the special conformal transformation
\begin{equation}\label{ special conformal transformation}
x\prime ^{\mu} =  \frac{(x ^{\mu} +a ^{\mu} x^2)}{(1 +2 a_{\nu}x^{\nu} +   a^2 x^2)}
 \end{equation}
for a certain D vector $a_{\nu}$.  which gives $\Omega^2 =\frac{1}{( 1 +2 a_{\mu}x^{\mu} +   a^2 x^2)^2} $
This result for $\Omega(x)^2$ can also be obtained by demanding that (\ref{Conformally transformed Minkowski}) satisfies the vacuum Einstein´s  Equation in $3+1$ dimensions \cite{Culetu}
or in any dimension \cite{braneworlds in string theories with dynamical tension}.
In summary, we have two solutions for the Einstein´s equations,
 $g^1_{\alpha \beta}=\eta_{\alpha \beta}$ and 
 
 \begin{equation}\label{ conformally transformed metric}
 g^2_{\alpha \beta}= \Omega^2\eta_{\alpha \beta} =\frac{1}{( 1 +2 a_{\mu}x^{\mu} +   a^2 x^2)^2} \eta_{\alpha \beta}
 \end{equation}
<
 These two spaces can be matched at the surfaces where $\Omega^2 = 1$, where
  $\Omega^2 =\frac{1}{( 1 +2 a_{\mu}x^{\mu} +  a^2 x^2)^2}$.
 We will consider the cases where  $a^2 \neq 0 $.
 Let us by consider the case where  $a^\mu$ is time like (for $a^\mu$ spacelike similar results are obtained), then without loosing generality we can take  $a_\mu = (A, 0, 0,...,0)$. The two solutions of $\Omega^2 = 1$ are , considering the cases $\Omega = 1$  and $\Omega = -1$, where $\Omega =\frac{1}{( 1 +2 a_{\mu}x^{\mu} +   a^2 x^2)} $, or what is the same, $ ( 1 +2 a_{\mu}x^{\mu} +   a^2 x^2) = 1 $ or  $ ( 1 +2 a_{\mu}x^{\mu} +   a^2 x^2) = -1 $ , The first case gives, 
 \begin{equation}\label{boundariesforBraneworld1}
2 a_{\mu}x^{\mu} +  a^2 x^2 = 0
\end{equation}
and the second case gives, 
\begin{equation}\label{boundariesforBraneworld2}
2 +2 a_{\mu}x^{\mu} +  a^2 x^2 = 0
\end{equation}
Then for  $a_\mu = (A, 0, 0,...,0)$, the above condition implies that, 
\begin{equation}\label{hyperbolic motion 1}
(2+ 2 a_{\mu}x^{\mu}  +  a^2 x^2) =
(2+ 2At + A^2(t^2-x^2)) =0
\end{equation}

This condition, if $A \neq 0$ implies then the more specific condition,
\begin{equation}\label{hyperbolic motion 1 expanded}  
 x^2_1  + x^2_2 + x^2_3.....+ x^2_{D-1}- (t+ \frac{1}{A})^2 = -\frac{1}{A^2}
\end{equation}

by a similar analysis, the other surface that satisfies the conformal transformation factor is $1$,  is given by (\ref{boundariesforBraneworld2}), which implies instead
\begin{equation}\label{hyperbolic motion 2}
 x^2_1  + x^2_2 + x^2_3.....+ x^2_{D-1}- (t+ \frac{1}{A})^2 = \frac{1}{A^2}
\end{equation}
 since (\ref{hyperbolic motion 1}) represents a surface that does not exist for all times, and it can represent propagation with speed greater than light,  we will restrict to the surface (\ref{hyperbolic motion 2}) which does exist for all times, and represents hyperbolic motion with time like propagation.
 
 Between the surface of most interest, that is the time like motion defined by equation  (\ref{hyperbolic motion 2}) and the tachyonic motion defined by eqs.
 (\ref{hyperbolic motion 1}) or
 (\ref{hyperbolic motion 1 expanded}) , there will be also the light cone surface,
 defined by $$1+ 2 a_{\mu}x^{\mu} +  a^2 x^2 = 0$$, which leads to
 $$x^2_1  + x^2_2 + x^2_3.....+ x^2_{D-1}- (t+ \frac{1}{A})^2 = 0$$,
 which does not define a surface suitable for matching the two spaces, but rather a surface where the conformal factor goes singular.

 By constructing our matching so that inside the radius defined by eq.  (\ref{hyperbolic motion 2}) we have just flat Minkowski space in Minkowski coordinates, we then eliminate the possibility of the tachyonic bubble and the  light cone pathology from the inside of the bubble, obtaining a well behaved space time. 
 
 In the next section we will look at the surface tension as given by the Israel matching conditions.
 
  \section{The Induced Space Time in the bubble is de Sitter}
 It is very interesting to notice that although we are matching two flat spaces, the induced metric in the bubble is a de Sitter space. This is indeed the case, if we recall the definition of de Sitter space as the induced metric in an hyperbolic surface moving in an embedding flat space see for example
 \cite{MISNERTHORNWHEELER}
de Sitter space can be defined in any dimension as a submanifold of a generalized Minkowski space of one higher dimension. Take Minkowski space in  $n+1$ the  $R^{1, n}$ space with the standard metric:
 \begin{equation}\label{Minkowski}
 ds_1^2 = \eta_{\alpha \beta} dx^{\alpha} dx^{\beta}
\end{equation}
then de Sitter space is the sub manifold described by the hyperboloid of one sheet
\begin{equation}\label{desitter}
 x^2_1  + x^2_2 + x^2_3 +  ...+.x^2_n - x^2_0 = \alpha^2
\end{equation}

where $ \eta_{\alpha \beta}$ is the standard Minkowski metric embedding space, with 
$ \eta_{00}= 1$, $ \eta_{0i}= 0 $ and $ \eta_{ij}= - \delta_{ij}$.
This is exactly our construction, since where the bubble lives it feels a Minkowski embedding space from both sides, since the conformal factor that relates the two sides equals one and eq.  (\ref{desitter}) is exactly of the form  (\ref{bubbleboundaryforBraneworld1bsurfacetension}), just that in our case the  (\ref{bubbleboundaryforBraneworld1bsurfacetension}) the origin of time is shifted for the hyperboloid. Notice that the embedding space is invariant under a shift of the origin oftime of course.
We can go deeper into the induced metric for the case of an embedding space of  $D+1=4$, defining $r= \sqrt{ x^2_1  + x^2_2 + x^2_3}$, then, taking the space (\ref{ conformally transformed metric}) outside, the standard Israel matching conditions \cite{Israel matching conditions} , transforming the two spaces to polar coordinates apply, where we have now,
 \begin{equation}\label{Minkowskipolar}
 ds_1^2 = - dt^2 + dr^2 + r^2 (d\theta^2 + sin(\theta)^2 d \phi^2 ) 
\end{equation}
and 
 \begin{equation}\label{Minkowskiconnformalpolar}
 ds_1^2 = \Omega(x)^2( - dt^2 + dr^2 + r^2 (d\theta^2 + sin(\theta)^2 d \phi^2 )) 
\end{equation}
so, we see that the the induced metric in the bubble is well defined and it results in the same induced metric, it does not matter if it is induced from the inside or from the outside because the conformal factor between them equals the identity at the juncture. The angular parts are the simplest to analyze, and in this case we obtain $g_{\theta \theta} = r^2$ and 
$g_{\phi \phi} = sin(\theta)^2  r^2$ in the inside, while $g_{\theta \theta} =\Omega(x)^2 r^2$ and
$g_{\phi \phi} = \Omega(x)^2 sin(\theta)^2 r^2 $  outside
so, while the metrics are continuous, and the induced metric at the surface is well defined at the surface since at that surface the conformal transformation $\Omega(x)^2 =1 $, 
The other metric component of the induced metric is obtained from the relation between the radial embedding coordinate $r$ and the time embedding coordinate $t$, for example, to be specific,  according to the relation (\ref{hyperbolic motion 2}), which means , defining  $r^2=x^2_1  + x^2_2 + x^2_3$, say in a three dimensional embedding space, that 
\begin{equation}\label{r and t for hyperbolic motion 2}
 r^2 = (t+ \frac{1}{A})^2 + \frac{1}{A^2}
\end{equation}
which implies that $$-dt^2 + dr^2 = -dt^2 \frac{1}{(A^2((t + \frac{1}{A})^2 +\frac{1}{A^2})}  = -  d\tau^2$$

which defines the proper time observed by a co-moving observer (that is an observer with $\theta$ and $\phi$  constant) in the bubble as a function of the embedding time. The induced space and induced metric in the bubble is then perfectly well defined from either side due to the fact that the conformal factor at the junction of the two spaces is one. 
The  above equation relating $t$ and  $\tau$ can of course be integrated, giving 
$$ At +1 = sinsh (A\tau) = $$
and using (\ref{r and t for hyperbolic motion 2}), we obtain $$ r^2 = \frac{1}{A^2}  cosh^2 (A\tau)$$, 
so, altogether the induced metric is a manifestly de Sitter $2+1$ metric
$$ds^2 = d\tau^2 + \frac{1}{A^2}  cosh^2 (A\tau) ( d\phi^2 + sin(\theta)^2  d\theta^2 ) $$
a very well known representation of a $2+1$ de Sitter space.
 \section{surface tension as given by the Israel matching conditions in four space-time dimensions}
The derivatives along the normal to the surface are not continuous  however, leading to a surface tension, for example from the discontinuity of the derivatives of the angular components of the metric along the normal to the surface, given by the Israel condition, 
\begin{equation}
    K^{+}_{\theta \theta}-K^{-}_{\theta \theta}=4 \pi G \sigma r^{2}
\end{equation}
where $ K^{+}_{\theta \theta}$ represents the $\theta \theta$ component of the extrinsic curvature in the outside region, which we take as the conformally transformed space and $K^{-}_{\theta \theta}$
is space inside the bubble, which we take as the Minkowski space in Minkowski coordinates,
\begin{equation}
  K^{+}_{\theta \theta}=\frac{1}{2}  \xi^{\mu} \partial_{\mu} (r^{2} \Omega^{2})
\end{equation}
\begin{equation}
  K^{-}_{\theta \theta}=\frac{1}{2}  \xi^{\mu} \partial_{\mu} (r^{2} )
\end{equation}
\begin{equation}
    K^{+}_{\theta \theta}-K^{-}_{\theta \theta}=\frac{1}{2}[\xi^{\mu} \partial_{\mu} (r^{2} \Omega^{2})-\xi^{\mu} \partial_{\mu} (r^{2} )]\Big \vert_{\Omega^{2}=1}= \frac{1}{2}\xi^{\mu}\Omega^{2} \partial_{\mu} (r^{2} )+\frac{1}{2}\xi^{\mu}r^{2} \partial_{\mu} ( \Omega^{2})-\frac{1}{2}\xi^{\mu} \partial_{\mu} (r^{2} )\Big \vert_{\Omega^{2}=1}
\end{equation}
\begin{equation}
  =  \frac{1}{2} \xi^{\mu}r^{2} \partial_{\mu} \Omega^{2}\Big \vert_{\Omega^{2}=1}=4\pi G \sigma r^{2}
\end{equation}

\begin{equation}
   \Omega^2 =\frac{1}{( 1 +2 a_{\mu}x^{\mu} +  a^2 x^2)^2}
\end{equation}
 then we have,
using the fact that the derivative of the conformal factor are of course orthogonal to the surface of conformal factor equal one, and then normalizing we find the expression for the normal to the surface, which leads to,  
 \begin{equation}
     \xi_{\mu}=\frac{\partial_{\mu}(\Omega^{2})}{\sqrt{|(\partial_{\mu}(\Omega^{2})(\partial_{\mu}(\Omega^{2})|}}   
 \end{equation}
 and then to the expression for the surface tension, 
 \begin{equation}
     \sigma=\frac{1}{8\pi}\sqrt{ | (\partial_{\mu}(\Omega^{2})(\partial_{\mu}(\Omega^{2})|}
 \end{equation}
 \begin{equation}
     \sigma=\frac{|A|}{4\pi}\sqrt { | 1+2a^{\mu}x_{\mu}+x^{2} | }
 \end{equation}
and using (\ref{boundariesforBraneworld1}) or (\ref{boundariesforBraneworld2}) we obtain 
$$ \sigma=\frac{|A|}{4\pi}$$
Notice however that the surface (\ref{hyperbolic motion 1 expanded}) moves with velocity bigger than light, while (\ref{hyperbolic motion 2}) moves with velocity less than light, so, we must choose 
(\ref{hyperbolic motion 2}) as the physical alternative. Using equation $ \sigma=\frac{|A|}{4\pi}$ we can express 
the  solution with time like motion of the surface separating the two vacuum solutions $\Omega^2 = 1$ in terms of the surface tension is,
\begin{equation}\label{bubbleboundaryforBraneworld1bsurfacetension}
 x^2_1  + x^2_2 + x^2_3.- (t+ \frac{1}{A})^2 = \frac{1}{16 \pi^2 \sigma^2}
\end{equation}
so in the limit of very large surface tensions we obtain that the minimum radius of the bubble becomes very small.
This shows similar features with results in \cite{universefromalmostnothing}, but here there are stronger,  since we 
do not just talk about  Universes out of almost empty space, but  Universes out of exactly empty space, already at the classical level.

\section{Discussion and Conclusions, consideration of Euclidean extension and the Quantum Creation Problem and the DE problem}
Here we want to have a preliminary discussion of a few subjects that should be investigated in more details in the future. 
We start with the quantum origin of the bubble universes study here. The surface $\Omega^2 = 1$, can be solved according to the option (\ref{hyperbolic motion 2}) which is the only one to appears to allow an Euclidean extension, by defining an Euclidean time
$(t_E + 1/A) =-i(t+ 1/A)$, which replaced into $\Omega^2 = 1$ gives the spherical euclidean region
\begin{equation}\label{bubbleboundaryforBraneworld1beuclidean}
 x^2_1  + x^2_2 + x^2_3.....+ x^2_{D-1}+ (t_E+ \frac{1}{A})^2 = \frac{1}{A^2}
\end{equation}
Defined in this way the induced euclidean extension, the euclidean version of a de Sitter space is a sphere.
The euclidean and real time regions defined above of the surface $\Omega^2 = 1$ can be discussed and whether they can be smoothly 
matched at $(t_E + 1/A) = (t+ 1/A) = 0 $. 
Notice that in the de Sitter  induced metric we found before, 
$$ds^2 =- d\tau^2 + \frac{1}{A^2}  cosh^2 (A\tau) ( d\phi^2 + sin(\theta)^2  d\theta^2 ) $$
can be converted into an euclidean metric by taking $\tau$ to be imaginary,$\tau_E= i \tau$ and then we get that $\tau_E$ becomes an angle, see \cite{Euclidean de Sitter space}, with period $2\pi/A$ and the euclidean metric being,  This periodicity is of course related to the Gibbons Hawking temperature of de Sitter space  \cite{GibbonsHawking}
$$ds^2 = d\tau_E^2 + \frac{1}{A^2}  cos^2 (A\tau_E) ( d\phi^2 + sin(\theta)^2  d\theta^2 ) $$
Notice that at $\tau_E = \tau =0$, we have $cos^2 (A\tau_E) = cosh^2(A\tau) =1$,
suggesting matching of euclidean and Minkowski signed metrics is possible at such point , the bouncing point in the Minkowski signed cosmology and the maximum radius of the Euclidean solution.
The combined, Euclidean plus  Minkowski signature
space must use therefore only half of the Euclidean de Sitter space, with the matching to its Minkowski signature corresponding space taking place at its equator.  

Notice however that the Israel formalism was not designed originally to match spaces with different signatures, although generalizations that allowed this have been proposed \cite{Time kink}, but it is not clear that an genuine quantum effect may be treated in a classical way, this deserves more investigation.
The quantum mechanical interpretation of the euclidean region is the tunneling region that eventually leads by matching to the creation of the baby universe. The first surface (\ref{hyperbolic motion 1 expanded}) cannot be extended to Euclidean space and it starts at $t=0$ from zero size in real space, but from our construction, that region has been eliminated even from the classical manifold and will not be discussed more.

A much more complex problem, but  related, is the nucleation of false vacuum bubbles   \cite{FARHIGUVENGUTH}, which is basically the extension to complex  spacetime of the classical solutions  
of false vacuum bubbles evolving in a Black Hole external space \cite{GUENDELMANGUTH} . 

 At the classical level, we have seen that the matching between two patches of flat space, one described by flat space in standard Minkowski coordinates and the other by flat space obtained after a special conformal transformation from the standard Minkowski coordinates. The two spaces can be made to match at the locations where the conformal transformation equals the identity, which gives two different possible motions for the surface separating the two spaces, we choose the solution which is manifestly time like, we can calculate the surface tension of the wall separating the two spaces and it turns out to be a constant.

 These type of situation, where we consider two flat spaces, one in Minkowski space and the other in flat space obtained after a special conformal transformation from the standard Minkowski coordinates and the surfaces where the conformal transformation is equal to one plays a special role in the case of string theories with dynamical tension, and these hypersurfaces are in higher dimensions in this case, these surfaces appear as surfaces where the string tensions go to infinity  causing an effective braneworld scenario
 \cite{braneworlds in string theories with dynamical tension}, 
 \cite{braneworlds in string theories with dynamical tension 2}. The approach here is a bit different, we do not rely on any ¨microscopic¨ theory, like string theory with dynamical tension, and just  simply match the two given spacetimes, obtaining nevertheless very similar situations in this way as well. 
 
 Notice that although we have started just from the matching of two flat spaces, the induced space in the bubble is a de Sitter type space time, which shows the possibilities of this braneworld scenario of explaining the Dark Energy without introducing by had a cosmological constant into the action, when consider ingthe higher dimensional cases, like $5$ dimensional  embedding space  with a $4$ dimensional hyper surface which will have the desired de Sitter induced space.

 \section{Acknowledgements}
 We want to thank The Bahamas Advanced Study Institute and Conferences (BASIC) 
 and Ben Gurion University for support of this project.
 
\end{document}